\documentstyle[12pt,epsf]{article}
\epsfverbosetrue
\def\figlabel#1{\xdef#1{\thefigure}}

\def\figalign#1#2#3#4#5#6{
\begin{figure}
\centerline{
\hbox to 2.5truein{\vtop{\hsize=2.5truein\epsfxsize=6cm
\centerline{\epsfbox{#1} }
\caption[]{#3}
\figlabel{#2} }}
\qquad\hbox to 2.5truein{\vtop{\hsize=2.5truein\epsfxsize=6cm
\centerline{\epsfbox{#4} }
\caption[]{#6}
\figlabel{#5} }} }
\end{figure} }

\def\be{\begin{equation}}
\def\ee{\end{equation}}
\def\bea{\begin{eqnarray}}
\def\eea{\end{eqnarray}}

\begin{document}
\begin{titlepage}
\begin{flushright} { ~}\vskip -1in CERN-TH/97-282\\ US-FT-32/97\\
hep-th/9710176\\ 
\end{flushright}
\vspace*{20pt}
\bigskip
\begin{center}
 {\Large Kontsevich Integral for Vassiliev Invariants \\ ~from
Chern-Simons Perturbation Theory in the Light-Cone Gauge}
\vskip 0.9truecm

{J. M. F. Labastida$^{a,b}$ and  Esther P\'erez$^{b}$}

\vspace{1pc}

{\em $^a$ Theory Division, CERN,\\
 CH-1211 Geneva 23, Switzerland.\\
 \bigskip
  $^b$ Departamento de F\'\i sica de Part\'\i culas,\\ Universidade de
Santiago de Compostela,\\ E-15706 Santiago de Compostela, Spain.\\}

\vspace{5pc}

{\large \bf Abstract}
\end{center} We analyse the structure of the perturbative series
expansion  of Chern-Simons gauge theory in the light-cone gauge. After
introducing a regularization prescription that entails the consideration
of framed knots, we present the general form of the vacuum expectation
value of a Wilson loop. The resulting expression turns out to give the
same framing dependence as the one obtained using non-perturbative
methods and perturbative methods in covariant gauges. It also contains
the Kontsevich integral for Vassiliev invariants of framed knots.

\vspace{6pc}

\begin{flushleft} { ~}\vskip -1in CERN-TH/97-282\\ October 1997\\
\end{flushleft}


\end{titlepage}

\def\theequation{\thesection.\arabic{equation}}

\def\del{{\delta^{\hbox{\sevenrm B}}}} \def\ex{{\hbox{\rm e}}}
\def\azb{A_{\bar z}} \def\az{A_z} \def\bzb{B_{\bar z}} \def\bz{B_z}
\def\czb{C_{\bar z}} \def\cz{C_z} \def\dzb{D_{\bar z}} \def\dz{D_z}
\def\im{{\hbox{\rm Im}}} \def\mod{{\hbox{\rm mod}}} \def\tr{{\hbox{\rm
Tr}}}
\def\ch{{\hbox{\rm ch}}} \def\imp{{\hbox{\sevenrm Im}}}
\def\trp{{\hbox{\sevenrm Tr}}} \def\vol{{\hbox{\rm Vol}}}
\def\rl{\Lambda_{\hbox{\sevenrm R}}} \def\wl{\Lambda_{\hbox{\sevenrm W}}}
\def\fc{{\cal F}_{k+\cox}} \def\vev{vacuum expectation value}
\def\nodiv{\mid{\hbox{\hskip-7.8pt/}}}
\def\ie{{\em i.e.}}

\def\np{Nucl. Phys.}
\def\pl{Phys. Lett.}
\def\prl{Phys. Rev. Lett.}
\def\pr{Phys. Rev.}
\def\ap{Ann. Phys.}
\def\cmp{Comm. Math. Phys.}
\def\ijmp{Int. J. Mod. Phys.}
\def\jmp{J. Math. Phys.}
\def\mpl{Mod. Phys. Lett.}
\def\inma{Invent. Math.}
\def\tam{Trans. Am. Math. Soc.}
\def\lmp{Lett. Math. Phys.}
\def\bams{Bull. AMS}
\def\am{Ann. of Math.}
\def\rmp{Rev. Mod. Phys.}
\def\jpsc{J. Phys. Soc. Jap.}
\def\topo{Topology}
\def\kjm{Kobe J. Math.}
\def\knot{Journal of Knot Theory and Its Ramifications}

\newcommand{\ZZ}{{\mbox{{\bf Z}}}}
\newcommand{\RR}{{\mbox{{\bf R}}}}
\newcommand{\MM}{{\mbox{{\bf M}}}}
\newcommand{\CC}{{\mbox{{\bf C}}}}
\newcommand{\RRs}{{\scriptstyle {\rm {\bf R}}}}
\newcommand{\MMs}{{\scriptstyle {\rm {\bf M}}}}
\newcommand{\CS}{{\scriptstyle {\rm CS}}}
\newcommand{\CSs}{{\scriptscriptstyle {\rm CS}}}
\newcommand{\beq}{\begin{equation}}
\newcommand{\eeq}{\end{equation}}
\newcommand{\bear}{\begin{eqnarray}}
\newcommand{\eear}{\end{eqnarray}}
\newcommand{\W}{{\cal W}}
\newcommand{\F}{{\cal F}}
\newcommand{\x}{{\cal O}}\newcommand{\LL}{{\cal L}}

\def\mani{{\cal M}}
\def\calo{{\cal O}}
\def\calb{{\cal B}}
\def\calw{{\cal W}}
\def\calz{{\cal Z}}
\def\cald{{\cal D}}
\def\calc{{\cal C}}
\def\to{\rightarrow}
\def\ele{{\hbox{\sevenrm L}}}
\def\ere{{\hbox{\sevenrm R}}}
\def\zb{{\bar z}}
\def\wb{{\bar w}}
\def\nodiv{\mid{\hbox{\hskip-7.8pt/}}}
\def\menos{\hbox{\hskip-2.9pt}}
\def\dr{\dot R_}
\def\drr{\dot r_}
\def\ds{\dot s_}
\def\da{\dot A_}\def\dga{\dot \gamma_}
\def\ga{\gamma_}
\def\dal{\dot\alpha_}
\def\al{\alpha_}
\def\cl{{\it closed}}
\def\cls{{\it closing}}
\def\vev{vacuum expectation value}
\def\tr{{\rm Tr}}
\def\to{\rightarrow}
\def\too{\longrightarrow}


\newfont{\namefont}{cmr10}
\newfont{\addfont}{cmti7 scaled 1440}
\newfont{\headfontb}{cmbx10 scaled 1728}
%

\section{Introduction}
\setcounter{equation}{0}

Chern-Simons gauge theory has provided a very useful tool to study
different aspects of the theory of knot and link invariants. Since its
formulation by Witten in 1988 \cite{csgt} it has been studied from both
perturbative and non-perturbative points of view. While non-perturbative
methods
\cite{csgt,nbos,torus,king,martin,kaul} have led to its connection to
polynomial invariants as the Jones polynomial \cite{jones} and its
generalizations \cite{homfly,kauffman,aku}, perturbative ones 
\cite{gmm,natan,alla,alts} have provided representations of Vassiliev
invariants.

One of the advantages of the Chern-Simons approach to Vassiliev invariants
is that it provides different representations of them because the
perturbative analysis of the theory can be carried out in different
gauges. Most of the perturbative analysis has been performed in covariant
gauges, mostly in the Landau gauge. The loop structure in this gauge have
been extensively analysed in
\cite{gmm,shift,piguet,cmartin}. Vacuum expectation values of Wilson loops
have also been studied in a variety of papers
\cite{gmm,natan,alla,alts,torusknots,factor}. These analyses lead  to a
representation of Vassiliev invariants for knots and links, which is
equivalent to the one proposed by Bott and Taubes in \cite{botttaubes}.

The perturbative Chern-Simons series expansion in non-covariant gauges has
not been studied so extensively. The loop structure has been analysed in
the light-cone gauge in \cite{leibrandt,sorella}. A first step into the
study of vacuum expectation values of Wilson loops was presented in
\cite{vande}, where a second-order analysis was carried out in the axial
gauge. No systematic study of the perturbative series expansion
corresponding to Wilson loops have been carried out in the light-cone
gauge. The main goal of this paper is to begin with this study. In the
process we find that the light-cone gauge leads to the Kontsevich integral
\cite{kont,tung} for Vassiliev invariants of framed knots.

Chern-Simons gauge theory was first studied in the light-cone gauge in
\cite{king}. That paper points out a close relation between the vacuum
expectation value of a Wilson line and the Knizhnik-Zamolodchikov
equations \cite{kzeqs}. This fact is used to prove that, for $SU(N)$ as
gauge group, vacuum expectation values of Wilson loops are related to the
HOMFLY polynomial \cite{homfly} for links. From a perturbative point of
view, vacuum expectation values of Wilson loops have been considered only
in \cite{cata}, where it is conjectured that the corresponding
perturbative series expansion is related to Kontsevich integral
\cite{kont} for Vassiliev invariants. In the present paper we prove this
conjecture.

Vacuum expectation values of Wilson loops are ill-defined in the
light-cone gauge because of the presence of singularities. To have a
well-defined perturbative expansion we introduce framed Wilson loops. The
fact that the right object to be studied in Chern-Simons gauge theory is a
framed Wilson loop has been known since the theory was first formulated in
\cite{csgt}. The reason to introduce a framing might result different in
each approach, but in general it is related to the presence of
singularities in the
$n$-point correlation functions at coincident points. This will indeed be
the case in the light-cone gauge. A version of Kontsevich integral for
Vassiliev invariants adapted to the case of framed knots and links was
presented in \cite{tung}. In this paper we will consider the case of
framed knots but a similar construction holds for framed links. We show
that the vacuum expectation value of a Wilson loop contains the Kontsevich
integrals introduced in \cite{tung}.

The paper is organized as follows. In sect. 2 we discuss the quantization
of Chern-Simons gauge theory in the light-cone gauge. In sect. 3 we
analyse the general features of the perturbative series expansion of
Wilson loops. In sect. 4, after introducing a regularization that
involves framed knots, we prove the finiteness of the perturbative series
expansion. In addition, we extract the framing dependence and show that
the perturbative series contains the Kontsevich integral for Vassiliev
invariants of framed knots. Finally, in sect. 5 we state our conclusions
and we discuss some open problems and future work.

\vfill
\eject

\section{Chern-Simons gauge theory in the light-cone gauge}
\setcounter{equation}{0}

In this section we introduce Chern-Simons gauge theory in the light-cone
gauge from a perturbation theory point of view. Let us consider a
semi-simple compact Lie group $G$ and a connection $A$. The action of  the
Chern-Simons theory over a three-dimensional Minkowski space
${\MM^3}$ is defined by the integral:
\beq S_\CS (A)={k\over 4\pi}\int_{\MMs^3} \tr  \Big(A\wedge dA +  {2\over
3} A\wedge A\wedge A\Big),
\label{action}
\eeq where Tr denotes the trace over the fundamental representation of
$G$, and $k$ is a real parameter. This action is invariant under gauge
transformation
\beq A_\mu \rightarrow h^{-1}A_\mu h + h^{-1} \partial_\mu h,
\label{gauge}
\eeq where $h$ is a map from $\MM^3$ to $G$, which is connected to the
identity map. For maps that are not connected to the identity,
(\ref{action}) transforms into itself plus a term of the form $2\pi k n$,
 $n$ being the winding number of the map. If $k$ is an integer, the
exponential of the action, $\exp(i S_\CS  )$, which is what enters the
functional integral, is invariant under both types of gauge
transformations.

Relative to the action (\ref{action}) and to the gauge transformation
(\ref{gauge}), we choose the following conventions. The gauge group
generators, $T^a$,  $a = 1 \dots {\hbox{\rm dim}}(G)$, will be taken
anti-Hermitian, satisfying
\beq [ T^a , T^b ]  =  - f_{abc} T^c,
\label{generators}
\eeq $f_{abc}$ being the structure constants of the group $G$. They are
normalized so that
\beq
\tr ( T^a T^b ) =  - {1 \over 2} \delta^{ab}, 
\label{generatorsdos}
\eeq in the fundamental representation.  A point in $\MM^3$ will be
denoted by coordinates $x^{\mu}$, with $\mu$  running from zero to two,
$\mu=2$ being the time coordinate.

In order to define the perturbative series expansion associated to the
action (\ref{action}) we must make a gauge choice. This choice is defined
by a gauge-fixing condition, which in our case will be the corresponding
to the light-cone gauge:
\beq n^{\mu} A_{\mu} = 0,
\label{light}
\eeq where $n^\mu$ is a constant vector satisfying $n^2=0$. The
corresponding gauge-fixing term to be added to the action (\ref{action})
is:
\beq S_{\scriptstyle {\rm gf}} = \int_{\MMs^3}  {\rm d}^3 x \tr ( d
n^{\mu} A_{\mu} + b n^{\mu} D_{\mu} c ),
\label{fixing}
\eeq where $d$ is an auxiliary field, and $c$ and $b$ are ghost fields;
$D_{\mu}$ stands for the covariant derivative, $D_\mu c = \partial_\mu c+
[A_\mu,c]$. In order to define the perturbative series expansion it is
convenient to  rescale the fields by $A \rightarrow g A$, where  
$g = \sqrt{4 \pi \over k}$. The quantum action, $S=S_\CS + S_{\scriptstyle
{\rm gf}},$ becomes:
\beq S =  -{1 \over 2}  \int_{\MMs^3} {\rm d}^3 x \epsilon^{\mu\nu\rho}  
\Big(A_{\mu}^a \partial_{\nu} A_{\rho}^a  - {g \over 3} f_{abc} 
A_{\mu}^a  A_{\nu}^b  A_{\rho}^c \Big)
 -  {1 \over 2} \int_{\MMs^3} {\rm d}^3 x \bigl( d^a A_{\mu}^a n^{\mu}  +
b^a  n^{\mu}  D_{\mu}^{ab} c^b,
\bigr)
\label{component}
\eeq where:
\beq D_{\mu}^{ab} = \partial_{\mu} \delta^{ab} - g f_{abc} A_{\mu}^c.
\label{covariant}
\eeq

Following \cite{king} we introduce light-cone coordinates:
\beq x^+ = x^1 + x^2,   \,\,\, \,\,\, x^- = x^1-x^2,
\label{coor}
\eeq and light-cone components for the gauge connection:
\beq A_+ = A_1 + A_2,   \,\,\, \,\,\, A_- = A_1-A_2.
\label{lcgc}
\eeq Choosing the vector $n^\mu$ as $(0,1,-1)$ the gauge-fixing condition
(\ref{light}) implies $A_-=0$. Notice that the gauge condition
(\ref{light}) does not fix the gauge completely. For example, in our
particular choice of $n^\mu$ we still have gauge invariance under gauge
transformations (\ref{gauge}), in which the gauge parameter $h$ depends
only on $x^0$ and $x^+$. The quantum action (\ref{component}) takes the
following form:
\beq S =    \int_{\MMs^3} {\rm d}^3 x  (A^a_+ \partial_- A^a_0 - A^a_0
\partial_- A^a_+  - b^a \partial_- c^a ).
\label{pino}
\eeq 

Actually, to consider this action as the quantum action of the theory
constitutes a rather simplified version of the full story. The
perturbative higher-loop analysis in axial-type gauges is a very delicate
issue, which requires to take into consideration some specific
prescription to regulate unphysical poles. Fortunately, this analysis has
been done for the case of Chern-Simons gauge theory in \cite{leibrandt}.
In these works it is shown that the effect of higher-loop contributions
is a shift of the parameter $k$ entering the Chern-Simons action
(\ref{action}) by the quantity $c_v$, which denotes the value of the
quadratic Casimir in the adjoint representation of the gauge group.
Though strictly speaking this has been proved at one loop, it is believed
that, as in the case of covariant gauges, it holds at any order in higher
loops. In this paper we will assume that higher-loop effects just account
for the shift of the parameter $k$ in (\ref{action}), and we will work
with (\ref{pino}) as the quantum action of the theory.

The quantum action (\ref{pino}) has three important properties. First, it
does not have derivatives in the transverse direction,  second, it is
quadratic
 in the fields, and, third,  the ghost fields are not coupled to the gauge
fields. This last property implies that the ghost fields can be integrated
out trivially. The second property implies that there are no interaction
vertices and therefore that all the correlation functions are determined
by the two-point correlation functions. Certainly, this property notably
simplifies the structure of the perturbative series expansion.

The two-point correlation functions corresponding to the gauge fields
entering (\ref{pino}) have been computed in \cite{king}. Their result is
more conveniently expressed after performing a Wick rotation to Euclidean
space $\RR\times \CC$.  A point in Euclidean space will be denoted as
$(t,z)$, where $z=x^1+i x^2$. After introducing $A_z = A_1 + i A_2$ and
$A_{\bar z} = A_1 - i A_2$ one finds \cite{king}:
\bear
\langle A_{\bar z}^a(x) A_m^b(x') \rangle &=& 0, \nonumber
\\  \langle A^a_m(x) A^b_n(x') \rangle &=&
 \delta^{ab} \epsilon_{mn} { 1 \over 2\pi i}  { \delta ( t - t'  )
\over z - z' },
\label{prop}
\eear with $m,n = \{ 0,z \}$, and $\epsilon_{mn}$ is antisymmetric with 
$\epsilon_{0z}=1$.

 \vfill
\eject

\section{Perturbative expansion of the Wilson loop}
\setcounter{equation}{0}

 Wilson loops are gauge-invariant operators of Chern-Simons gauge theory
labelled by a loop $C$ embedded in ${\MM^3}$ and a  representation $R$ of
the gauge group $G$. They are defined by the holonomy  along the loop $C$
of the gauge connection $A$:
\beq
\W_R(C,G)=\left[{\hbox{\rm P}}_R \exp g \oint A \right],
\label{wilsonloop}
\eeq where ${\hbox{\rm P}}_R$ denotes that the integral is path-ordered
and that
$A$ must be considered in the representation $R$ of $G$. In Chern-Simons
gauge theory one considers vacuum expectation values of products of
Wilson-line operators:
\bear &&\langle \W_{R_1}(C_1,G)\W_{R_2}(C_2,G)\dots\W_{R_n}(C_n,G)\rangle 
\nonumber\\   && \,\,\,\,\,\,\,\,\, = {1\over Z_k}
\int [DA] \W_{R_1}(C_1,G)\W_{R_2}(C_2,G)\dots\ 
\W_{R_n}(C_n,G){\rm e}^{iS_k(A)}, 
\label{vev}
\eear where $Z_k$ is the partition function of the theory:
\begin{equation} Z_k=\int [DA]\, {\rm e}^{iS_{CS}(A)}.
\label{parfun}
\end{equation}

\begin{figure}
\centerline{\hskip.4in \epsffile{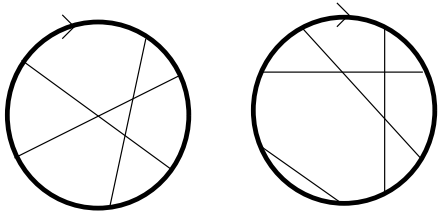}}
\caption{Examples of Feynman diagrams.}
\label{fdiag}
\end{figure}

As shown in \cite{csgt} the quantity (\ref{vev}) is a link invariant 
associated to a coloured $n$-component link whose $j$-component $C_j$
carries  the representation $R_j$ of the gauge group $G$. In this paper we
will consider only the \vev\ of a single  loop $\langle \W_{R}(C,G)
\rangle$. This quantity can be expressed as a perturbative series
expansion in the coupling constant $g$, which is the result of evaluating
all the corresponding Feynman diagrams. The structure of these diagrams is
obtained using standard field theory methods. A typical diagram of order
$i$ consists of a solid thick circle representing the oriented path $C$,
with $i$ propagators attached to it in a certain order (see some examples
in fig. \ref{fdiag}). A term in the sum of the perturbative series  may be
regarded as constructed out of two Feynman rules: the one corresponding to
the propagator (\ref{prop}):
\beq D_{mn}^{ab}(x-x') = \delta^{ab} \epsilon_{mn} { 1 \over 2\pi i}  {
\delta (t - t' ) \over z - z' }, 
\label{maspropa}
\eeq while all other components of $D_{\mu\nu}$ vanish; and the one
corresponding to the vertex between the end-point of a propagator and the
oriented path $C$:
\beq V_{i}^{j\, \mu a}(x) = g (T^a_{(R)})_i^j \int {\rm d} x^\mu.
\label{frules}
\eeq Notice that, as discussed in the previous section, there are not
three-vertices in the light-cone gauge. The Feynman rules are depicted in
fig. 2.

The power series expansion corresponding to $\langle \W_{R}(C,G) \rangle$
can be written as \cite{alla}:    
\beq
\langle \W_R(C,G)  \rangle={\rm dim}\,R\sum_{i=0}^{\infty}\sum_{j=1}^{d_i}
\alpha_{ij}(C)\,r_{ij}(R)\,x^i,
\label{general}
\eeq where  $x=ig^2/2$ is the expansion parameter. The quantities
$\alpha_{ij}(C)$, or geometrical factors, are combinations
 of path integrals of some kernels along the loop $C$,  and the $r_{ij}$
are  traces of products of generators of the Lie algebra associated to the
gauge  group $G$. The  index $i$ corresponds to the order in perturbation
theory, and 
$j$ labels independent contributions at a given order, $d_i$  being the
number  of these at order $i$. In (\ref{general}) ${\rm dim}\,R$ denotes
the dimension of the representation $R$. Notice the convention
$\alpha_{01}(C) = 1$. For a given order in perturbation theory,
$\{r_{ij}\}_{\{j= 1 \dots d_i\}}$ represents a basis of independent group
factors. 

\begin{figure}
\centerline{\hskip.4in \epsffile{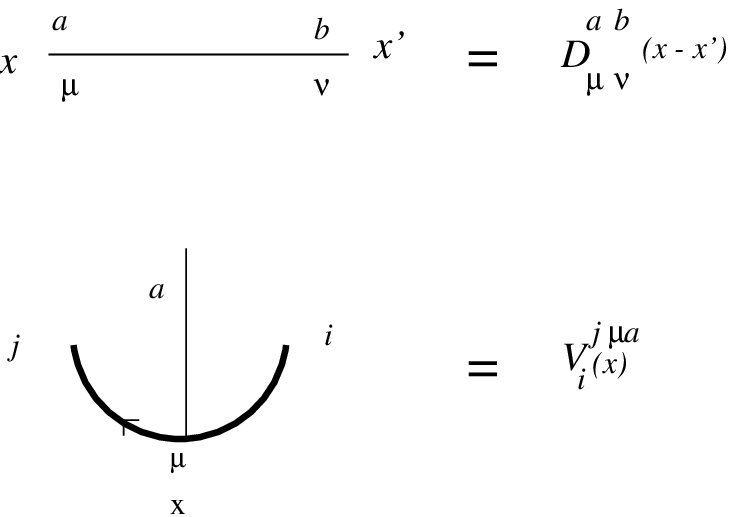}}
\caption{Feynman rules.}
\label{propa}
\end{figure}

\begin{figure}
\centerline{\hskip.4in \epsffile{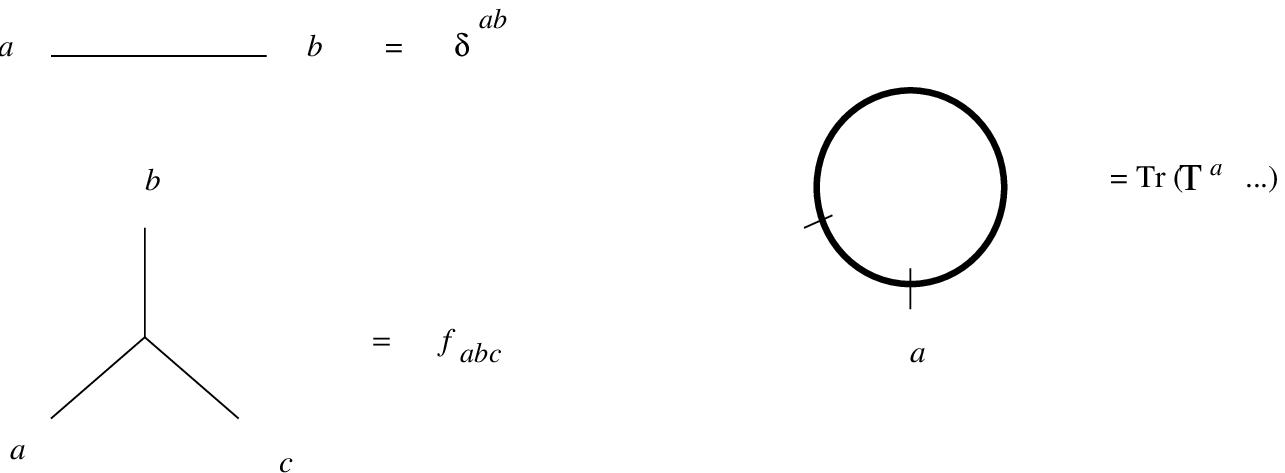}}
\caption{Group-theoretical rules.}
\label{groupr}
\end{figure}

The choice of basis in (\ref{general}) is not unique. As shown in
\cite{factor}, there is a special type of  basis, called canonical, for
which the series (\ref{general}) satisfies a factorization theorem. Let us
introduce some notation  to recall the definition of a canonical basis. A
group factor is an element in the centre of the universal enveloping
algebra of $G$, its general form being a trace over products of generators
and  structure constants with all indices contracted. It may also be
represented in terms of diagrams that look like  Feynman diagrams  in
which propagators and three-vertices are attached to a circle in a certain
order, which represents the trace. In this way one considers three
group-theoretical Feynman rules as the ones depicted in fig.
\ref{groupr}.  A canonical basis consists of an independent set of group
factors  made out of connected  diagrams or products of non-overlapping
connected subdiagrams (see \cite{factor} for more details). 

The factorization theorem \cite{factor} states that (\ref{general})  may
be written as an exponentiation over the elements of a canonical basis,
which are not product of connected subdiagrams. These elements are called
primitive and (\ref{general}) takes the form:
\beq
\langle   \W_R(C,G) \rangle = {\rm dim} R \,
\exp\left\{ \sum_{i=0}^{\infty}\sum_{j=1}^{\hat{d}_i}
\alpha^c_{ij}(C)r_{ij}^c(R)x^i \right\}.
\label{primitive}
\eeq In this expression the superindex $c$ denotes the primitive connected
diagrams. The quantities  $\alpha^c_{ij}(C)$ are their corresponding
geometrical factors, which in general are linear combinations of the
$\alpha_{ij}(C)$ entering (\ref{general}). In (\ref{primitive}), 
$\hat{d}_i$ denotes the number of independent primitive group factors at
order $i$. For a more detailed discussion about the properties of 
(\ref{primitive}) we refer the reader to \cite{factor}. Here we will
recall only a few facts. In the series expansion (\ref{general}) of
$\langle   \W_R(C,G) \rangle$  there is a class of diagrams that contain
at least one collapsible propagator. A propagator is called collapsible
whenever its two legs are attached to two points in the oriented path $C$,
which can be considered as the end-points of a part of $C$ in which no
other propagator is attached. Notice that in the right-hand side of
(\ref{primitive}) all these diagrams have factorized out as the
exponentiation of the diagram of order 1, its geometrical factor being 
$\alpha^c_{11} (C)$. The perturbative analysis carried out in the Landau
gauge in \cite{alla} demonstrated that if one chooses a canonical basis
all the dependence on the framing is contained in $\alpha^c_{11} (C)$. In
this paper we will show that, as expected, this also holds when the theory
is analysed in the light-cone gauge.

As shown in \cite{bilin,barnatan}, each term of the perturbative series
expansion (\ref{general}) is a Vassiliev invariant. The quantities 
$\alpha^c_{ij}(C)$, $i>1$, in the exponential of (\ref{primitive})
represent a particular choice of a basis of primitive ones. The analysis
of the perturbative series expansion of the theory in the light-cone gauge
will provide integral expressions for these invariants. Actually, in order
to make contact with the Kontsevich integral we are not going to construct
the integral expression for the primitive elements
$\alpha^c_{ij}(C)$, $i>1$. We will not make the  choice of a canonical
basis except in our discussion on the framing dependence. Instead, we will
analyse the perturbative series expansion as it results from the
application of the Feynman rules without selecting a particular basis of
group factors.  As will be shown in the next section, this will lead us
to Kontsevich integral for Vassiliev invariants.

\begin{figure}
\centerline{\hskip.4in \epsffile{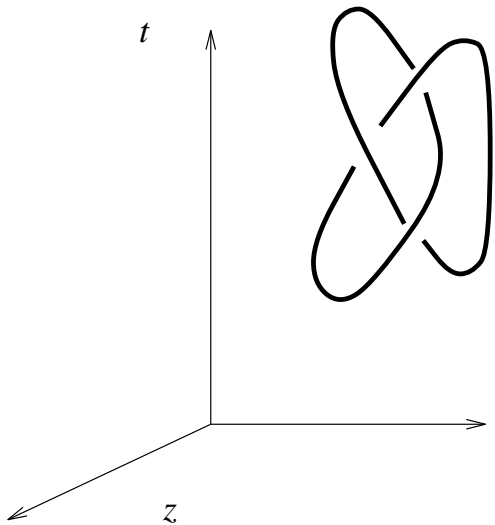}}
\caption{Example of a Morse knot.}
\label{exparam}
\end{figure}

Before entering into the analysis of the perturbative series expansion
corresponding to a Wilson loop, we must discuss the potential problems
that might be encountered because of the particular form of the
gauge-field propagator (\ref{maspropa}). This propagator is singular when
its two end-points coincide. Actually, it is particularly singular in this
situation, because both the numerator and the denominator lead to
divergences. This fact tells us that, as in other gauges, one must
consider Wilson loops with oriented paths $C$ with no self-intersections.
In other words, one must consider knots. However, contrary to covariant
gauges, in the light-cone gauge we have two special kinds of
singularities, there may be situations in which only one, the numerator
or the denominator, leads to a divergence.  In order to avoid
singularities from the numerators one is forced to avoid paths with
sections in which $t$, the first component of a generic point $(t,z)$ in
$\RR\times\CC$, is constant. This constraint, together with the fact that
one is only allowed to consider the non-self-intersecting paths, implies
that one must consider paths $C$, which correspond to Morse knots. A Morse
knot in
$\RR\times\CC$ is a knot in which $t$ is a Morse function on it. A Morse
knot is characterized by $2n$ extrema, half of them maxima, and the other
half minima. An example of a Morse knot is depicted in fig.
\ref{exparam}. 

The third potential problem due to the structure of (\ref{maspropa}) comes
from situations in which the two end-points of the propagator are close to
one of the extrema of a Morse knot since  the denominator then vanishes.
To solve this problem we will introduce a regularization procedure based
on the introduction of a framing for the knot. The resulting invariants
will correspond to invariants of framed knots.

\vfill
\eject

\section{The Kontsevich integral for framed knots}
\setcounter{equation}{0}

In this section we will apply the Feynman rules derived in the previous
section to  construct the perturbative series expansion corresponding to a
Wilson loop for a Morse knot. As argued before, this requires the
introduction of a regularization procedure; this will be achieved by
considering framed knots. In the first part of this section we will
describe this procedure in detail. In three subsequent subsections we
will  first prove the finiteness of the resulting perturbative series
expansion, then we will extract the framing dependence, and, finally, we
will obtain the Kontsevich integral for framed knots.

Let us consider a framed oriented Morse knot $K$ on $\RR\times\CC$. The
framing of $K$ is defined by a vector $m^\mu(K)$ normal to $K$ such that
along the coordinates $(t(a),z(a))$, $a\in S^1$, which define $K$, we have
another set of coordinates $(t(a),z(a))+\varepsilon (m^0(t,z),m^z(t,z))$,
which define a companion knot $K_\varepsilon$. For small $\varepsilon$,
$K_\varepsilon$ is also a Morse knot, which does not intersect $K$ and
becomes $K$ in the limit $\varepsilon\rightarrow 0$. We assign to 
$K_\varepsilon$ the orientation that is naturally inherited from $K$. The
framing is characterized by the vector $m^\mu(K)$ or, equivalently, by the
companion knot $K_\varepsilon$. It is natural to associate to each choice
of framing the integer number that corresponds to the linking number,
$l$, between $K$ and $K_\epsilon$.

Our regularization of the vacuum expectation value of a Wilson loop is
based on the replacement of the propagator (\ref{maspropa}), which is
attached to two points on $K$, by a propagator attached to a point in $K$
and another in $K_\varepsilon$. As it is described below, there exists a
precise way to carry out this replacement. Certainly, the resulting
regularized integrals entering the power series expansion are finite for
Morse knots. We have to show that they remain also finite in the limit
$\varepsilon\rightarrow 0$. Before doing this, let us give full details
of how the propagator replacement is carried out.

Let us assume that the Morse knot $K$ possesses $2n$ extrema. There are
$2n$ curves $k^i$, $i=1,\dots,2n$, joining  the different maxima and
minima of $K$. For each curve $k^i$ there is a one-to-one correspondence
between points on
$k^i$ and the values that  the variable $t$ takes.  The Morse knot can
therefore be regarded as a complex multivalued function of the variable
$t$ with $2n$ components, each one corresponding to a curve $k^i$, which
is completely labelled by the values that the complex variable $z$ takes
in
$k^i$ as a function of $t$. One can think of two different
parametrizations of $K$, the previous one: $(t(a),z(a))$, $a\in S^1$, and
a new one: $z_i(t)$,
$t\in I^i$, $i=1,\dots,2n$, where $I^i=[t_i^-,t_i^+]$ is the segment of
$\RR$ whose end-points are the values that the coordinate $t$ takes at the
two extrema joined by the curve $k^i$. A similar analysis can be done for
the companion knot $K_\varepsilon$. It also possesses $2n$ extrema and
$2n$ curves $k^i_\varepsilon$, parametrized by $z'_i(t)$, $t\in
I^i_\varepsilon$, $i=1,\dots,2n$.

It is convenient to trade the integrations along the $S^1$ parameter $a$
of
$K$ by integrations over the height parameter $t$. Let us consider a
propagator attached to the curves $k^i$ and $k^j$. Using (\ref{maspropa}),
it is easy to show that for such a propagator:
\beq d x^\mu d x^\nu \langle A^a_\mu(x) A^b_\nu(x') \rangle {1\over 2\pi
i}
\rightarrow {1\over 2\pi i} d t_i d t_j \delta(t_i-t_j) p_{ij} { \dot
z_i(t_i) - \dot z_j(t_j) \over z_i(t_i) - z_j(t_j)},
\label{trading}
\eeq where:
\beq p_{ij} = 
\left\{  \begin{array}{lll} 
 1 & {\rm if} \,\, k_i \,\, {\rm and } \,\, k_j \,\, {\rm have} \,\, {\rm
the} \,\, {\rm same} \,\, {\rm orientation,}  &  \\  -1 &  {\rm if} \,\, 
k_i \,\, {\rm and } \,\, k_j \,\, {\rm have}  \,\, {\rm opposite} \,\,
{\rm orientations.} &  \end{array}
\right. 
\label{sign}
\eeq

The propagators (\ref{trading}) may appear attached to two points lying
on different curves $k^i$ and $k^j$, $i\neq j$, or to two points lying on
the same curve $k^i$. In the second case the regularization consists of
just replacing one of the points on $k^i$ of the propagator by a point on
$k^i_\varepsilon$. Since the delta function in the propagator
(\ref{trading}) implies that its two end-points must be at the same
height, there is no ambiguity in doing this. Notice, however, that in
this case, since we have a path-ordered integration,  when evaluating the
delta  function in (\ref{trading}) we  must take into account the
appearance of a factor $1/2$. In the first case, $i\neq j$, one uses the
same procedure, namely one of the end-points is attached to the curve
corresponding to the companion knot. However,  there are now two possible
ways of doing this. We will use the following prescription: we will add
both possibilities and multiply the result by a factor $1/2$. For the same
reasons as in the second case, this last step is well defined. Notice that
in this case there is no additional factor $1/2$ after the evaluation of
the delta functions, since the variables of integration in each curve
are  now free. 

The regularization prescription leads to the following formulae for the
propagator (\ref{trading}). For $i=j$:
\beq {1\over 2 \pi i} {1\over 2} d s {\dot z_i(s) - \dot z_i'(s) \over
                                       z_i(s) - z_i'(s) },
\label{laura}
\eeq while for $i\neq j$:
\beq {1\over 2 \pi i} {1\over 2} d s \Big(  {\dot z_i(s) - \dot z_j'(s)
\over z_i(s) - z_j'(s) }+ {\dot z_i'(s) - \dot z_j(s) \over z_i'(s) -
z_j(s) } \Big) p_{ij}.
\label{martin}
\eeq

\begin{figure}
\centerline{\hskip.4in \epsffile{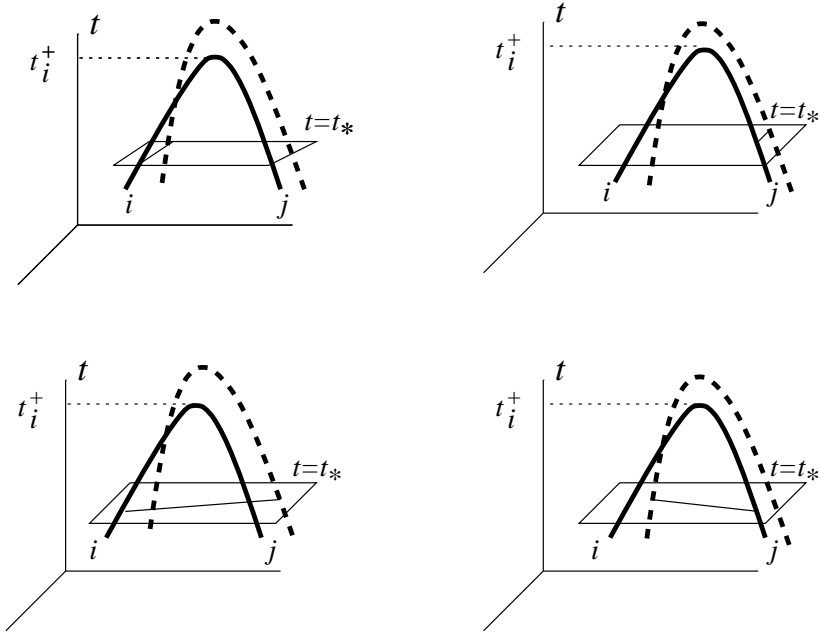}}
\caption{Generic divergent contributions.}
\label{diver}
\end{figure}

\subsection{Finiteness}

\begin{figure}
\centerline{\hskip.4in \epsffile{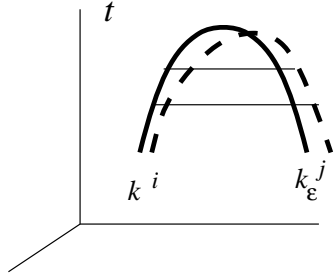}}
\caption{Higher-order divergent contributions.}
\label{dosrayas}
\end{figure}

We will now prove that our prescription is finite in the limit
$\varepsilon\rightarrow 0$. There are four sources of singularities. All 
originate when the two end-points of a propagator are near an extremum. We
will describe the situation for the case in which this extremum is a
maximum, but it will become clear that a similar analysis can be carried
out in the case in which it corresponds to a minimum. The four
configurations that lead to divergences are depicted in fig. \ref{diver}.
We have drawn only the part of $K$ around the maximum. We will assume that
the rest of the diagram is the same in the four cases.  Let us compute the
four contributions after integrating from $t_*$ to $t^+_i$. Using
(\ref{laura}) and (\ref{martin}) one finds, after adding them up:
\beq {1\over 2\pi i}{1\over 2}\log{(z_i(s)-z_i'(s))(z_j(s)-z_j'(s))\over
                    (z_i(s)-z_j'(s))(z_i'(s)-z_j(s))}
\Bigg|_{s=t_*}^{s=t^+_i}
\label{pera}
\eeq The most dangerous contribution is the one coming from the upper
limit. However, since $|z_i(t^+_i)-z_i'(t^+_i)|=
|z_j(t^+_i)-z_i'(t^+_i)|=\varepsilon$, and
$|z_i(t^+_i)-z_j'(t^+_i)|= |z_j(t^+_i)-z_j'(t^+_i)|=\varepsilon$, one
finds a cancellation of divergences in the limit $\varepsilon\rightarrow
0$ and therefore the contribution is finite. One also encounters
divergences from the lower limit of (\ref{pera}) in the 
$\varepsilon\rightarrow 0$ limit. However, these divergences cancel
against the ones originated after integration under the height $t_*$. One
is therefore left with a term of the form:
\beq -{1\over 2\pi i}{1\over 2} \log
\big((z_i(t_*)-z_j'(t_*))(z_i'(t_*)-z_j(t_*))\big),
\label{sobra}
\eeq which has to be taken into account when integrating over $t_*$.
Notice that (\ref{sobra}) is finite in the limit $\varepsilon\rightarrow
0$ for values of $t_*$ away from $t^+_i$, but it diverges when $t_*$
approaches $t^+_i$. However, this divergence is too soft and it does not
generate singularities in the integration, except in situations of the
kind depicted in fig. \ref{dosrayas}. All the contributions of this type
will be treated below when analysing the factorization of the framing
dependence, and it will be shown that all possible divergences cancel out.
We can therefore affirm that the perturbative series expansion
corresponding to a Morse knot is finite in the limit
$\varepsilon\rightarrow 0$.

\subsection{The framing contribution}

In this subsection we are going to compute the lowest-order contribution
to the perturbative series expansion of the vacuum expectation value of a
Wilson loop. Using the factorization theorem discussed in sect. 3, we will
obtain the full dependence on the framing. Our result can be stated very
simply: if $K$ is a framed Morse knot:
\beq
\langle {\cal W}_R(K,G)\rangle = 
\ex^{2\pi i l h} \langle {\cal W}_R(K,G)\rangle',
\label{limon}
\eeq where $h=\tr(T^a_{(R)}T^a_{(R)})/k$, $l$ is the linking number
between $K$ and its companion knot $K_\varepsilon$, and $\langle {\cal
W}_R(K,G)\rangle'$ is a framing-independent quantity. This result agrees
with the one found non-perturbatively \cite{csgt}, and in covariant
gauges \cite{alla}.

\begin{figure}
\centerline{\hskip.4in \epsffile{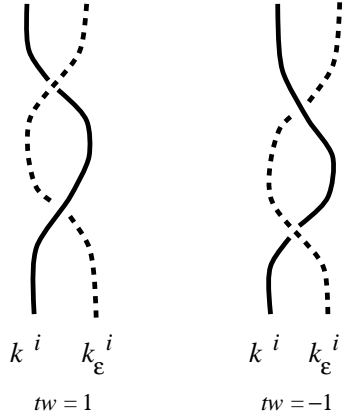}}
\caption{Basic contributions to the twist of a framed knot.}
\label{twist}
\end{figure}

\begin{figure}
\centerline{\hskip.4in \epsffile{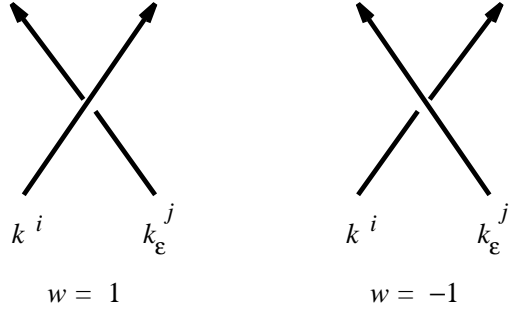}}
\caption{Basic contributions to the writhe.}
\label{writhe}
\end{figure}

The lowest-order contribution to the perturbative series expansion has the
form:
\beq ig^2 {1\over 2\pi i}{1\over 2}\tr(T^a_{(R)}T^a_{(R)})
\sum_{i,j=1}^{2n} 
\int^{{\scriptstyle{\rm min}}\{t_i^+,t_j^+\}}_{{\scriptstyle{\rm
max}}\{t_i^-,t_j^-\}} d s {\dot z_i(s) - \dot z_j'(s)\over z_i(s) -
z_j'(s)}p_{ij}.
\label{sandia}
\eeq Notice that, owing to the prescription (\ref{martin}), the sum is
over all possible values of $i$ and $j$ and not only for $i\leq j$.  The
factor
$i$ in front of this expression is due to the fact that one is considering
the functional integral of $i$ times the action (\ref{component}). In
general, in the perturbative series expansion, one must include a factor
$i$ for each power of $g^2$. According to our previous arguments, the
integral (\ref{sandia}) is certainly  finite in the limit
$\varepsilon\rightarrow 0$. Actually it can be computed very easily. The
key observation is that an integral of the form
\beq
\int_{t_*}^{t_*'} d s {\dot z_i(s) - \dot z_j'(s)\over z_i(s) - z_j'(s)}
\label{laintegral}
\eeq develops an imaginary part, which counts the number of times that the
curve
$k^j_\varepsilon$ winds around the curve $k^i$, times $2\pi$. The real
parts just cancel with each other, following a mechanism similar to the
one described in our proof of finiteness. After summing over all the
values of $i$, the contribution from the integral in (\ref{sandia}) for
$i=j$ is just 
$2\pi i$ times the twist $tw$ of the band made by the two knots $K$ and
$K_\varepsilon$, \ie\ the number of times that the band twists  around
itself. The basic contributions to $tw$ are shown in fig. \ref{twist}. The
analysis of the contributions from $i\neq j$ is very similar. The real
parts cancel among themselves and some of the real contributions from the
case $i=j$, as described in the proof of finiteness.  The imaginary parts
now count the number of times that the curve $k^i$ twists around the curve
$k^j$, times $2\pi$. Certainly, for this counting one can take the limit
$\varepsilon\rightarrow 0$. If the Morse knot $K$ is viewed in such a way
that its projection in the plane corresponding to Im $z =0$ contains only
single crossings, one can compute the contribution just assigning values
$\pm 1/2$ to the two types of crossings shown in fig. \ref{writhe}. The
result of performing the integral in (\ref{sandia}) is precisely $2\pi i$
times the writhe $w$ of the knot $K$ for that given view. Notice that
although we get factors of modulus $1/2$ from each crossing, they are
counted twice in (\ref{sandia}). Thus, the full contribution from
(\ref{sandia})  is:
\beq ig^2{1\over 2}\tr(T^a_{(R)}T^a_{(R)}) (w+tw)= 2 \pi i h l
\label{final}
\eeq since $l=w+tw$. In obtaining (\ref{final}) we have used 
$g^2=4\pi/k$. Recall that $h=\tr(T^a_{(R)}T^a_{(R)})/k$. Notice that,
although $w$ and $tw$ depend on the view of the knot $K$, their
combination $l$, the linking number between $K$ and
$K_\varepsilon$,  is independent of it.

\begin{figure}
\centerline{\hskip.4in \epsffile{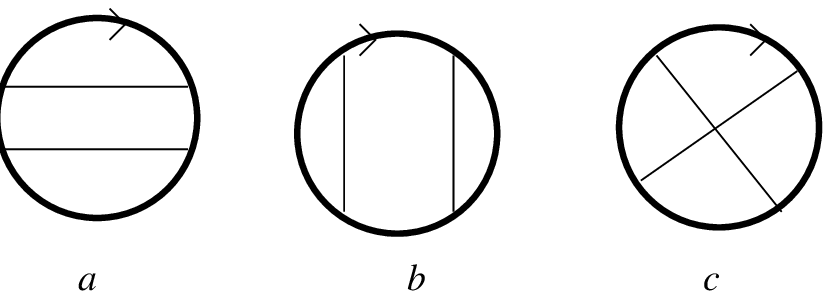}}
\caption{Feynman diagrams contributing at second order.}
\label{diafac}
\end{figure}

As shown in previous works \cite{alla,factor}, the contribution found at
first order exponentiates if one considers a canonical basis. Let us
describe the reasons for this exponentiation by analysing what occurs at
next order in perturbation theory. At second order one encounters the two
group factors corresponding to the Feynman diagrams shown in fig.
\ref{diafac}. Diagrams {\it a} and {\it b} share the same group factor:
\beq 
\tr(T^a_{(R)}T^a_{(R)}T^b_{(R)}T^b_{(R)}),
\label{factoruno}
\eeq while diagram {\it c} possesses a different one:
\beq 
\tr(T^a_{(R)}T^b_{(R)}T^a_{(R)}T^b_{(R)}).
\label{factordos}
\eeq If one decomposes this group factor in terms of the first one and a
new group factor using the commutation relations (\ref{generators}):
\beq 
\tr(T^a_{(R)}T^b_{(R)}T^a_{(R)}T^b_{(R)})=
\tr(T^a_{(R)}T^b_{(R)}T^b_{(R)}T^a_{(R)})
-f_{abc}\tr(T^a_{(R)}T^b_{(R)}T^c_{(R)}),
\label{factortres}
\eeq the full sum of the geometrical terms multiplying the first group
factor can be written as $1/2$ times a product of integrations over
contributions coming from a single propagator. It precisely gives the
second-order term of the exponential of (\ref{final}). A similar
rearrangement can be done at any order. Notice that in order to extract
the terms building the exponential of (\ref{final}) one modifies the group
factors that remain in the rest. This mechanism is encoded in the
factorization theorem \cite{factor}. Recall that in our analysis of
finiteness we postponed the discussion on the cancellation of divergences
from diagrams as the one in fig. \ref{dosrayas}. Now is the moment to
discuss that issue. These diagrams are indeed the ones having group
factors as (\ref{factoruno}). Together with the ones coming after
rearranging group factors as done in (\ref{factortres}), after adding them
up, one obtains  an expression that can be written as a product of the
contribution from the first order. Since that first-order contribution is
finite, the contribution at any arbitrary higher order is also finite.
Notice also that this argument shows that all the dependence on the
framing is contained in the exponential factor and therefore we can affirm
that (\ref{limon}) holds. Indeed all the potential dependence on the
framing must come from diagrams that lead to divergences. As we have
seen, these diagrams only contribute to the exponential factor in
(\ref{limon}). This behaviour is entirely similar to the one occurring in
the Landau gauge \cite{alla}.

\subsection{The Kontsevich integral}

In this subsection we will prove that the perturbative series expansion
for the vacuum expectation value of the Wilson loop contains the
Kontsevich integral for framed knots as presented in \cite{tung}. 

Let us begin by writing  all the contributions to a given order $m$. To
carry this out we must consider all possible ways of connecting  $2m$
points on the Morse knot by $m$ propagators, following the regularization
prescription described in the previous section, \ie\ with one point of
the propagator attached to $K$ and the other to its companion knot
$K_\varepsilon$, and then path-order integrating. The path-order
integration can be split into a sum such that in each term enters a
path-ordered integration along $2m$ curves among the set $k^i$, 
$k^i_\varepsilon$, $i=1,\dots,2n$. This set of curves builds  the Morse
knot under consideration.  A given term in this sum might contain
propagators joining  $k^i$ and $k^i_\varepsilon$. In this case one must
introduce a factor $1/2$ as explained in our discussion leading to
(\ref{laura}). The contributions coming from propagators joining $k^i$
and $k^j_\varepsilon$, with $i\neq j$, have also a factor $1/2$ due to the
double counting, as explained in the discussion of eq. (\ref{martin}).
Accordingly, propagators joining different curves must be replaced by
$1/2$ the sum of their two possible choices of attaching their
end-points. 

\begin{figure}
\centerline{\hskip.4in \epsffile{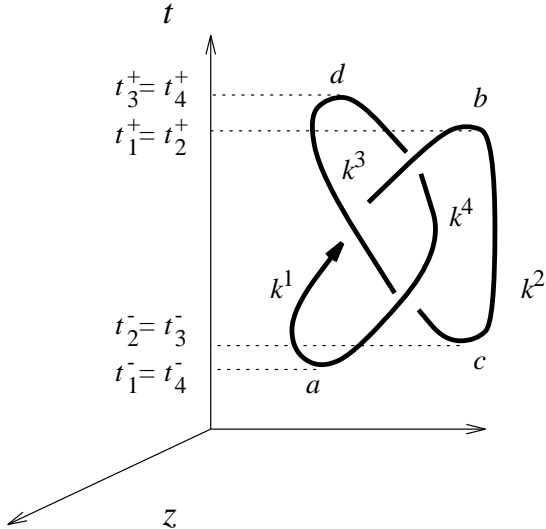}}
\caption{Labelled trefoil knot.}
\label{trlabel}
\end{figure}

To each rearrangement of the $m$ propagators corresponds a group factor.
These are easily obtained using the group-theoretical Feynman rules shown
in fig. \ref{groupr}. To fix ideas we will present in detail the
second-order contribution, $m=2$, for a particular group factor. For $m=2$
one must take into consideration the two group factors (\ref{factoruno})
and (\ref{factordos}). As discussed above, the group factor
(\ref{factoruno}) is associated to the framing dependence. We will
analyse the contribution corresponding to the group factor
(\ref{factordos}). This contribution is of the form
\bear && (ig)^2 \int_0^1 d s_1 \int_{s_1}^1 d s_2 \int_{s_2}^1 d s_3
\int_{s_3}^1 d s_4 \dot x^{\mu_1}(s_1) \dot x^{\mu_2}(s_2)
\dot x^{\mu_3}(s_3) \dot x^{\mu_4}(s_4) \nonumber \\ &&{\hskip-1cm}
\langle A_{\mu_1}^{a_1}\big( x(s_1) \big)
        A_{\mu_3}^{a_3}\big( x(s_3) \big) \rangle
\langle A_{\mu_2}^{a_2}\big( x(s_2) \big)
        A_{\mu_4}^{a_4}\big( x(s_4) \big) \rangle
\tr ( T^{a_1} T^{a_2} T^{a_3} T^{a_4}). 
\label{chamo}
\eear We will now write more explicitly this multiple integral taking into
account the form of the propagator ({\ref{maspropa}). The delta function
in this propagator imposes very strong restrictions on the possible
contributions. Its presence implies that the only non-vanishing
configurations are those in which the two end-points of each propagator
are at the same height. To be more concrete let us consider the
computation of (\ref{chamo}) for the trefoil knot shown in fig.
\ref{trlabel}. This knot is made out of four curves, $k^1$, $k^2$, $k^3$
and $k^4$, whose end-points are the four critical points $a$, $b$, $c$ and
$d$. The heights of these critical points are:
\bear
 a &\rightarrow & t_1^- = t_4^-, \nonumber \\
 b &\rightarrow & t_1^+ = t_2^+, \nonumber \\
 c &\rightarrow & t_2^- = t_3^-,  \\
 d &\rightarrow & t_3^+ = t_4^+. \nonumber 
\label{abcd}
\eear They are depicted in fig. \ref{trlabel}. 

\begin{figure}
\centerline{\hskip.4in \epsffile{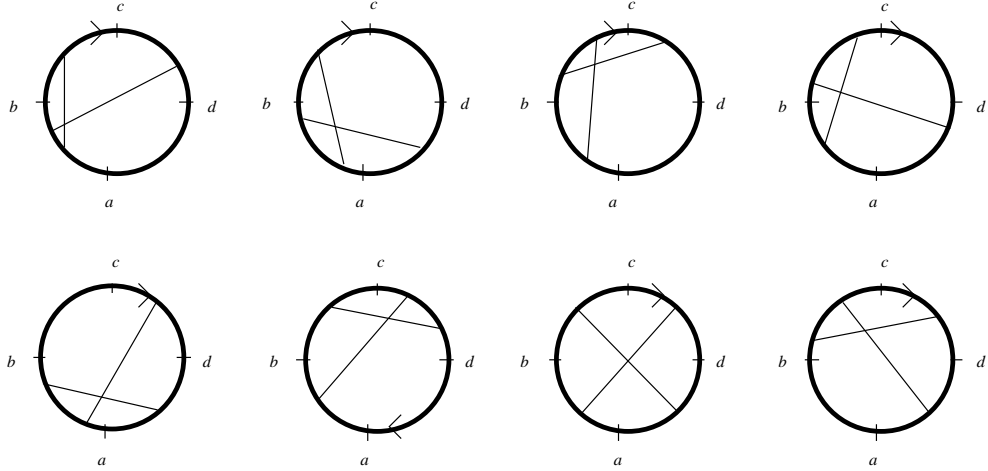}}
\caption{Contributions to a Morse knot made out of four curves.}
\label{balon}
\end{figure}

\begin{figure}
\centerline{\hskip.4in \epsffile{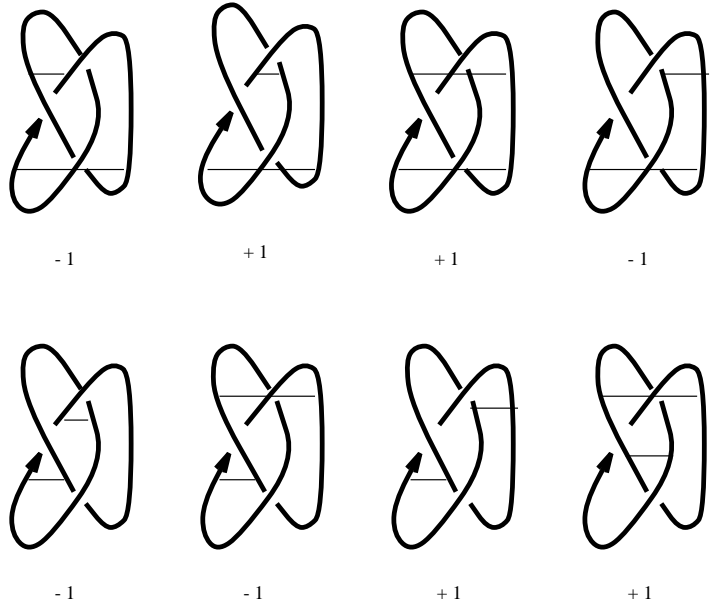}}
\caption{Contributions corresponding to the trefoil knot.}
\label{onkn}
\end{figure}

To obtain all the contributions we will divide  in four parts the circle
that represents the knot in fig. \ref{trlabel}. Then we will join these
parts by lines representing the propagators, taking into account the
ordering of the four points to which they are attached. This ordering and
the delta function in the height imply that no line can have its two
end-points  attached to the same part. They also imply that there are no
contributions in which two end-points of different lines are attached to
one part and the other two to another part. The only possibilities are
shown in fig.
\ref{balon}. There is a total of eight contributions. Notice that this
result is general for any Morse knot made out of four curves. The
contributions  are easily depicted on the knot itself as shown in fig.
\ref{onkn}. For each contribution, one must compute a sign, which is the
product of the $p_{ij}$ in (\ref{sign}). The resulting signs are
displayed in fig. \ref{onkn}. To be more explicit, let us write, for
example, the integral associated to the first contribution. It takes the
form
\bear && (ig)^2 {1\over (2\pi i)^2} {1\over {2^2}}
\int_{t_2^-<s_1<s_2<t_1^+} d s_1 d s_2 \nonumber\\
&&{\hskip-1.7cm}\Bigg({\dot z_1(s_1)-\dot z_2'(s_1) \over z_1(s_1) -
z_2'(s_1)} +{\dot z_1'(s_1)-\dot z_2(s_1) \over z_1'(s_1) -
z_2(s_1)}\Bigg)
\Bigg({\dot z_3(s_2)-\dot z_1'(s_2) \over z_3(s_2) - z_1'(s_2)} +{\dot
z_3'(s_2)-\dot z_1(s_2) \over z_3'(s_2) - z_1(s_2)}\Bigg).\nonumber\\
\label{greta}
\eear Similar expressions can be easily written for the rest of the
contributions depicted in fig. \ref{onkn}. The data entering in the double
integral (\ref{greta}) are shown in fig. \ref{calex}. Notice that this
integral is not divergent if we take the limit $\varepsilon
\rightarrow 0$ before performing the integration. This feature is common
to all the contributions corresponding to the group factor under
consideration. As shown in our discussion on finiteness, only the
contributions related to framing are potentially divergent. One could
therefore remove in (\ref{greta}) the terms with primes and the factor
$1/2^2$. The integral to be computed takes the form:
\be (ig)^2 {1\over (2\pi i)^2} \int_{t_2^-<s_1<s_2<t_1^+} d s_1 d s_2
{\dot z_1(s_1)-\dot z_2(s_1) \over z_1(s_1) - z_2(s_1)}{\dot z_3(s_2)-\dot
z_1(s_2) \over z_3(s_2) - z_1(s_2)}.
\label{meza}
\ee One of the two integrations can easily be performed, leading to:
\be (ig)^2 {1\over (2\pi i)^2} \int_{t_2^-<s_2<t_1^+} d s_2 \log
\Bigg({z_1(s_2)-z_2(s_2) \over z_1(t_2^-) - z_2(t^-_2)}\Bigg) {\dot
z_3(s_2)-\dot z_1(s_2) \over z_3(s_2) - z_1(s_2)}.
\label{yamoko}
\ee Notice that, as argued before, this integral is finite. Although $z_1$
and $z_2$ get close to each other when $s_2\rightarrow t_1^+$, the
singularity in the integrand, being logarithmic, is too mild to lead to a
divergent result. Expressions similar to (\ref{yamoko}) can easily be
obtained for the rest of the contributions depicted in fig. \ref{onkn}.

\begin{figure}
\centerline{\hskip.4in \epsffile{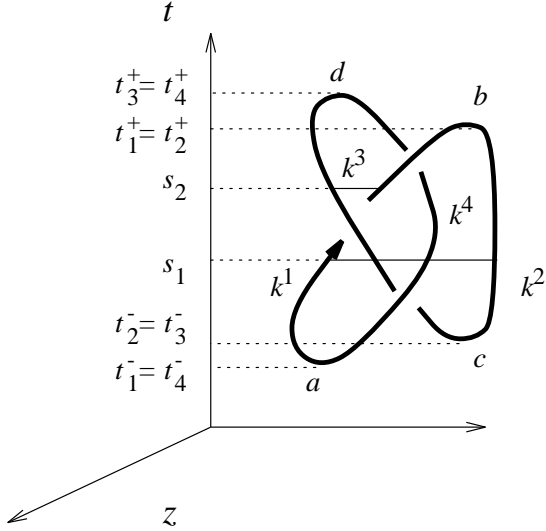}}
\caption{Details on the labeling of one of the contributions.}
\label{calex}
\end{figure}

We are now in  a position to write the form of the general contribution
originated from the Feynman rules of the theory. Notice that the most
significant fact of our previous discussion is the presence of the delta
function in the height of the propagator (\ref{maspropa}). It implies that
the only non-vanishing configurations of the propagators are those in
which their two end-points have the same height; in other words, only
contributions in which the line representing the propagator is horizontal
do not vanish. This observation allows us to rearrange the contributions
to the perturbative series expansion in the following way. Consider all
possible pairings
$\{z_i(s),z_j'(s)\}$ of curves $k^i$ and $k^j_\varepsilon$,
$i,j=1,\dots,2n$, where $2n$ is the number of extrema of the Morse knot
under consideration.   A contribution at order $m$ in perturbation theory
will involve a path-ordered integral in the heights
$s_1<\dots<s_r<\dots<s_m$ of a product of $m$ propagators of the form
(\ref{laura}) and (\ref{martin}):
\beq
\prod_{r=1}^m {d z_{i_r}(s_r) - d z_{j_r}'(s_r)\over z_{i_r}(s_r) -
z_{j_r}'(s_r)}.
\label{elproducto}
\eeq This product is characterized by a set of $m$ ordered pairings, each
one labelled by a pair of numbers $(i_r,j_r)$ with $r=1,\dots,m$. We will
denote an ordered pairing of $m$ propagators generically by $P_m$. One
must take into account all possible ordered pairings, \ie\ one must sum
over all the possible $P_m$. The group factor that corresponds to each
ordered pairing $P_m$ is simply obtained by placing the group generators
at the end-points of the propagators and taking the trace of the product,
which results after traveling along the knot. The resulting group factor
will be denoted by $R(P_m)$.

Another ingredient in (\ref{martin}) that must be taken into account is
the factor $p_{ij}$. For each pairing $P_m=\{(i_r,j_r),r=1,\dots,m\}$
there will be a contribution from their product. Certainly, the result
will be a sign that will depend on the ordered pairing $P_m$. We will
denote such a product by:
\beq s(P_m) = \prod_{r=1}^m  p_{i_r j_r}.
\label{elsigno}
\eeq

We are now in a position to write the full expression  for the
contribution to the perturbative series expansion at order $m$. It takes
the form:
\beq (ig^2)^m\Big({1\over 2\pi i}\Big)^m {1\over 2^m}\sum_{P_m}
\int_{t_{P_m}^-<t_1<\dots <t_r< \dots<t_m<  t_{P_m}^+} {\hskip-1cm} s(P_m)
\prod_{r=1}^m {d z_{i_r}(t_r) - d z_{j_r}'(t_r)\over z_{i_r}(t_r) -
z_{j_r}'(t_r)} R(P_m),
\label{eltung}
\eeq where $t_{P_m}^+$ and $t_{P_m}^-$ are highest and lowest heights,
which can be reached by the last and first propagators of a given ordered
pairing $P_m$. This expression corresponds to the Kontsevich integral for
framed knots as presented in \cite{tung}. 

\vfill
\eject

\section{Conclusions and open problems}

In this paper we have presented an analysis of Chern-Simons gauge theory
in the light-cone gauge from a perturbative point of view. As it became
clear in our discussion of sect. 3, only vacuum expectation values
associated to Morse knots are suitable for calculation in the light-cone
gauge, at least if no additional regularization to the one considered in
the paper is introduced. We have shown that the regularization
prescription leads to the consideration of framed knots, and that the
vacuum expectation value possesses  a dependence on the framing that
agrees with the ones appearing in other gauges and in the
non-perturbative analysis of the theory. We have also shown that the
perturbative series expansion contains the Kontsevich integral for framed
knots.

Our results, however,  demonstrate that something has been missed in the
perturbative series expansion. Indeed, contrary to what  is obtained in
other gauges or in non-perturbative approaches, according to our result
(\ref{eltung}), the vacuum expectation value corresponding to the unknot
shown in fig. \ref{unknot}  carrying a representation $R$ of the gauge
group is just dim$R$ (in the trivial framing). In other words, all the
contributions in (\ref{eltung}) vanish and one is left with the
zeroth-order contribution, which is just dim$R$. Actually, the expression
(\ref{eltung}) does not lead to knot invariants. It is well known
\cite{kont} that the Kontsevich integral is only invariant under
deformations of the knot which preserve the number of critical points. To
obtain a truly invariant quantity one must take into account a correction
to the sum of the contributions (\ref{eltung}). If we denote by
$Z_m(K,K_\varepsilon)$ the contribution shown in (\ref{eltung}) (order
$m$), the quantity that leads to a knot invariant is \cite{kont}:
\be {{\rm dim}R + \sum_{m=1}^\infty Z_m(K,K_\varepsilon) \over 
\Big(1 + {1\over {\rm dim}R}\sum_{m=1}^\infty 
Z_m(U,U_\varepsilon)\Big)^{n\over 2}}
\label{groot}
\ee where $U$ and $U_\varepsilon$ are the unknots shown in fig.
\ref{elotro}, and $n$ is the number of critical points of the knot $K$.
The coefficients of the powers of $g$ in this expression are Vassiliev
invariants. 

\begin{figure}
\centerline{\hskip.4in \epsffile{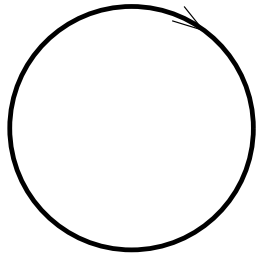}}
\caption{Unknot with two critical points.}
\label{unknot}
\end{figure}

\begin{figure}
\centerline{\hskip.4in \epsffile{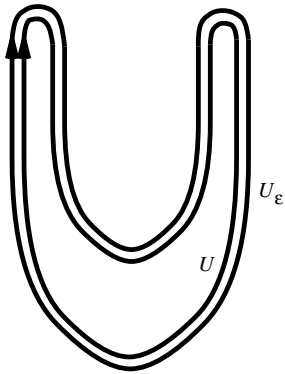}}
\caption{Unknots $U$ and $U_\varepsilon$ with four critical points.}
\label{elotro}
\end{figure}

The fact that the perturbative series expansion corresponding to the terms
(\ref{eltung}) is not invariant under all kinds of deformations of the
knot is something that one could have expected. As discussed in sect. 3,
the perturbative construction fails for knots that are not of the Morse
type. In a deformation of the knot in which the number of critical points
is changed, one has to consider at some intermediate step a knot that is
not of the Morse type. The analysis presented here is not applicable to
that situation and  it is therefore consistent to find that the final
expression for the perturbative expansion is not invariant under those
types of deformations. A very interesting problem is to understand, from
the point of view of Chern-Simons gauge theory, why the denominator of
(\ref{groot}) has to be included there. Presumably, this would reconcile
the results obtained here with the ones from covariant gauges and from
non-perturbative approaches. This and other related issues are at present
under investigation.

\vspace{4 mm}

\begin{center} {\bf Acknowledgements}
\end{center}

\vspace{4 mm}

We would like to thank M. Alvarez for helpful discussions.  This work was
supported in part by DGICYT under grant PB93-0344.

\end{document}